\documentclass{IEEEtran}
\usepackage{array}
\usepackage{amssymb}
\usepackage{balance}
\usepackage{booktabs}
\usepackage{comment}

\usepackage{enumitem}
\usepackage{etoolbox}
\usepackage[colorlinks,urlcolor=blue,linkcolor=blue,citecolor=blue]{hyperref}
\usepackage[capitalise,nameinlink]{cleveref} 
\usepackage{multirow}
\usepackage{natbib}
\usepackage{orcidlink}
\usepackage{pifont}
  \newcommand{\cmark}{\ding{51}}%
\usepackage{siunitx}
\usepackage{soul}
\usepackage{stfloats}
\usepackage{tabularx}
\usepackage[table]{xcolor}
\usepackage{xparse}
\usepackage{xspace}

\expandafter\def\expandafter\UrlBreaks\expandafter{\UrlBreaks\do\/\do\*\do\-\do\~\do\'\do\"\do\-}

\newcommand{\agenticAI}{agentic~AI\xspace}
\newcommand{\AgenticAI}{Agentic~AI\xspace}

\newcommand{\arxivonly}[1]{#1}
\begin{document}

\title{Rising From the Ashes: How Agentic AI is Unblocking Challenges in Cybersecurity}

\author{\IEEEauthorblockN{Gabriela F. Ciocarlie$^1$, Kathrin Grosse$^2$, Somesh Jha$^3$,  \\ Daryna Oliynyk$^4$, Andrew Paverd$^5$, Christian Wressnegger$^6$ \\}
\IEEEauthorblockA{$^1$Stevens Institute of Technology, USA $^2$Independent Researcher, Italy $^3$University of Wisconsin-Madison, USA $^4$CDL AsTra, Faculty of Computer Science, University of Vienna, Austria \\ $^5$Microsoft Security Response Center (MSRC), UK \\ $^6$KASTEL Security Research Labs, Karlsruhe Institute of Technology (KIT), Germany}}

\markboth{}{}

\maketitle
\begin{abstract}
Security remains a high-cost challenge, with many problems historically deemed inefficient to address or effectively unsolvable. A significant number of these problems stem from labor-intensive tasks that create bottlenecks in defensive approaches.  \AgenticAI has the potential to alleviate these bottlenecks by directly ingesting and reasoning over natural language or code, thereby expanding the scope of feasible defenses. In this paper, we map open security problems to emergent agentic AI capabilities. To illustrate this potential, we examine 16 case studies, including supply chain analysis, highlighting how agentic AI may benefit defenders. 
\end{abstract}

\section{Introduction}
The annual cost of cybersecurity incidents is estimated to range from \SIrange{0.001}{9}{\percent} of global GDP~\cite{cobos2024review}.
These high costs reflect that security is far from solved. Over the past decades there have been many proposals for techniques and approaches that could prevent attacks or solve contemporary cybersecurity challenges.
However, many of these have been held back from reaching their full potential because they are inherently labor-intensive, knowledge-intensive, slow, or error-prone.
For example, although we now have memory-safe programming languages, we still have large amounts of legacy code because the human effort required to rewrite this in a new language exceeds the benefit.
We have the tools to formalize and verify standards and protocols, but very few have actually been verified.
We have techniques for attribution and threat intelligence, but these are largely unused 
due to scarce expertise. 

Recent developments in AI have enabled direct, language-based communication with AI systems\arxivonly{~\cite{radford2019language}}. Fine-tuning these approaches has shown promising capabilities in code analysis\arxivonly{~\cite{jiang2026survey}}, potentially processing large amounts of data, and integration of specific, general, or interdisciplinary knowledge.
In addition, AI systems have become more autonomous as they are integrated directly with tools\arxivonly{~\cite{zhou2024webarena,acharya2025agentic}}.
In contrast to humans, AI agents may have the cognitive stamina to scan and analyze large code bases and reason about them.
\AgenticAI may thus not only enable mere automation of security workflows, but also redefine how we conceive, build, and deploy security tools at scale. 

In this paper, we explore how \agenticAI can catalyze reviving capabilities or exploring entirely new approaches to cybersecurity.
To this end, we first summarize \textbf{five} challenges (C1-C5) that have impeded the use of security techniques, and then outline \textbf{five} potential \agenticAI capabilities (A1-A5) that could ameliorate these challenges.

To illustrate, consider the example of \emph{feasibility studies} that check whether an attack applies to realistic targets, whether a proposed defense holds against a class of threats, or whether a new mechanism can be prototyped end-to-end at all.
Although the question they aim to answer may be simple, such studies can be expensive and slow to carry out, contradicting the \emph{``fail fast, learn fast''} principle.
Based on its natural language understanding, information processing, and generalization capabilities, \AgenticAI can solve previous challenges.
An \agenticAI can produce code, conduct experiments, and reason about outcomes, and can thus perform feasibility studies at substantially reduced cost and on shorter timescales, addressing previous challenges of excessive human effort, lowering previously high latency, providing potentially scarce expertise, and providing task-specific solutions. 

We argue that the application of \agenticAI provides a potential general solution to security challenges.
To this end, we  elaborate on \textbf{16} case studies spanning the domains of systems security, software security, attacks and defenses, and cyber threat intelligence (CTI).
However, the promise of a universal solution must be matched with rigor:
agents entrusted with security-critical tasks must themselves be secure, offering strong guarantees that their autonomy does not introduce unforeseen vulnerabilities.

\begin{table*}[t]
  \centering
  \caption{Mapping between AI agent capabilities and the security challenges they help address.}
  \label{tab:capabilities}
  \renewcommand{\arraystretch}{1.3}
\rowcolors{2}{gray!8}{white} 

\begin{tabularx}{\textwidth}{
    X
    >{\centering\arraybackslash}p{1.5cm}
    >{\centering\arraybackslash}p{1.5cm}
    >{\centering\arraybackslash}p{1.5cm}
    >{\centering\arraybackslash}p{1.5cm}
    >{\centering\arraybackslash}p{1.5cm}
}
\toprule
\rowcolor{gray!20}
\textbf{Potential AI Capability} & \textbf{Excessive human effort} & \textbf{Unattainably low latency} & \textbf{Scarce expertise} & \textbf{Task-agnostic solution} & \textbf{Problem complexity} \\
\midrule
Natural language understanding       & \cmark &            & \cmark & \cmark & \cmark \\
Processing large amounts of information & \cmark & \cmark &            &            & \cmark \\
Cognitive stamina                     & \cmark &            &            &            & \cmark \\
Generalization and replicability       & \cmark & \cmark & \cmark & \cmark &            \\
Multi-disciplinary knowledge           &            &            & \cmark & \cmark & \cmark \\
\bottomrule
\end{tabularx}
\end{table*}

\section{PREVIOUS CHALLENGES}
In this section, we highlight five factors previously posed as challenges to the widespread adoption or deployment of various security mechanisms, and for each of them, we discuss how modern \agenticAI systems can help to change the status quo.
\begin{enumerate}[label=\bf(C\arabic*),wide=0em]
\item\textbf{Requires excessive human effort:} 
The task requires significant human time and effort, and the cost of doing this is likely to exceed the value obtained. 
For example, writing a complete formal specification for a widely-used security protocol may be sufficiently valuable to justify the effort, but doing the same for a rarely-used specialized protocol may not be justifiable.

\item\textbf{Requires unattainably low latency:} 
The task must be completed within a bounded time, 
but humans are generally unable to complete the task within this time.
For example, to be effective against an active attack, a self-healing system may need to be adapted/healed faster than a human is capable. 

\item\textbf{Requires scarce/unavailable expertise:} 
The task requires specialized expertise or skills, but there are not enough people with the requisite skills or expertise to carry out the task at scale.
For example, finding vulnerabilities in deployed software systems requires specific skills, but there is a scarcity of people with these skills relative to the demand.

\item\textbf{Task-specific effort:}
A solution for one instance of the task does not generalize to other instances of the task, thus making widespread deployment or adoption very costly. 
For example, translating legacy code into memory-safe programming languages requires a specific solution for each code base. 

\item\textbf{Problem complexity:}
The task is inherently complex, with exceptionally many constraints or aspects to be considered. This could imply that solving the task requires excessive human effort or scarce expertise, but could also extend into the realm of tasks that are beyond human ability. 
\end{enumerate}

\begin{table*}[b!]
  \centering\vspace*{-2mm}
  \caption{Defensive ideas and their limitations}
  \label{tab:case_studies}
  \renewcommand{\arraystretch}{1.25}
\setlength{\tabcolsep}{8pt}

\rowcolors{2}{gray!8}{white} 

\begin{tabularx}{\textwidth}{
    >{\centering\arraybackslash}p{0.2cm}
    >{\raggedright\arraybackslash}X
    >{\centering\arraybackslash}p{1.5cm}
    >{\centering\arraybackslash}p{1.5cm}
    >{\centering\arraybackslash}p{1.5cm}
    >{\centering\arraybackslash}p{1.5cm}
    >{\centering\arraybackslash}p{1.5cm}
}
\toprule
\rowcolor{gray!20}
 & \textbf{Defensive idea} & \textbf{Excessive human effort} & \textbf{Unattainably low latency} & \textbf{Scarce expertise} & \textbf{Task-specific solution} & \textbf{Problem complexity} \\
\midrule
& \hyperref[sec:instrumentation]{Optimizing Instrumentation for Security}  & \cmark     & \cmark & \cmark  & \cmark & \cmark \\
& \hyperref[sec:moving-target]{Moving-Target Defense \& SW~Diversification}  &  & \cmark & \cmark & \cmark  &  \cmark \\
& \hyperref[sec:formalization-of-sec-policies]{Formalization of Security Standards}  & \cmark &        & \cmark & \cmark &  \\
& \hyperref[sec:supply-chain-analysis]{Supply Chain Analysis} &  \cmark  &  &        & \cmark  & \cmark \\
& \hyperref[sec:self-healing]{Self-Healing Systems}  &        & \cmark &        & \cmark & \cmark \\
\multirow{-6}{*}{\rotatebox{90}{Systems Security}} & \hyperref[sec:system-security]{Customization of System Security}  &  &        &  & \cmark  & \cmark \\
\midrule 
& \hyperref[sec:vulnerability-discovery]{Vulnerability Discovery} & \cmark &  & \cmark & \cmark & \cmark \\
& \hyperref[sec:exploit-generation]{Exploit Generation} & \cmark &        & \cmark &  \cmark & \cmark \\
& \hyperref[sec:translation-of-legacy-code]{Translation to Memory-Safe Language}  &  \cmark &  & \cmark & \cmark &  \\
& \hyperref[sec:domain-optimization]{\mbox{Domain-Specific Code Optimization}}  &  &  & \cmark  &  \cmark &        \\
\multirow{-5.5}{*}{\rotatebox{90}{Software Security~}} & \hyperref[sec:code-ip]{Code IP Protection}  &  &        &        & \cmark & \cmark \\
\midrule 
& \hyperref[sec:attribution]{Attribution and Threat Intelligence} & \cmark &        & \cmark & \cmark & \cmark \\
\multirow{-2}{*}{\rotatebox{90}{CTI}}
& \hyperref[sec:provenance-analysis-forensics]{Provenance Analysis for Forensics} & \cmark &  &        &  &  \cmark \\
\midrule 
& \hyperref[sec:insider-threat-detection]{Insider Threat Detection} & \cmark  & \cmark &  & \cmark & \cmark \\
& \hyperref[sec:provenance-analysis-monitoring]{Provenance Analysis \& RT Monitoring} &  \cmark & \cmark &        &   & \cmark \\
\multirow{-3.3}{*}{\rotatebox{90}{Attack Det.}} & \hyperref[sec:dynamic-honeypots]{Adaptive honeypots} &        & \cmark &  & \cmark &  \\
\bottomrule
\end{tabularx}
\end{table*}

\section{POTENTIAL AI CAPABILITIES}

Having described the challenges that have kept security ideas from scaling, we now turn to the capabilities of AI agents that could address them. 
We present five capabilities that current agents have demonstrated to varying degrees and that are improving as the technology matures. \Cref{tab:capabilities} provides an overview of how each capability addresses the challenges outlined in the previous section.
\begin{enumerate}[label=\bf(A\arabic*),wide=0em]
    \item \textbf{Natural language understanding:} Unlike conventional software, AI agents can process and reason in natural language, while still using ordinary programming languages for implementation. This capability enables operation at a higher level of abstraction, which reduces the manual effort and the expertise required to solve complex problems and supports more general, system-independent solutions. 
    
    \item \textbf{Processing large amounts of information:} AI~agents are capable of processing large amounts of material at once and reasoning over it jointly, rather than examining it piece by piece like humans. This both reduces the human effort required to work through large problems, helps address inherently complex tasks where the relevant context is too broad to hold in view all at once, and allows a conclusion to be reached quickly enough for tasks that must be completed within a tight-time bound.

    \item \textbf{Cognitive stamina:} AI agents deliver consistent judgments across similar tasks, consistently applying the same level of scrutiny without the fatigue or attention drift that affect human analysts. This sustained cognitive `agentic stamina' reduces the burden of prolonged analysis and is especially valuable for complex tasks, where maintaining uniform standards across many cases or constraints is challenging for~humans.

    \item \textbf{Generalization and replicability:} Building and validating an AI agent for a specific task is a one-time investment; once complete, the agent can be cloned and deployed in parallel across many systems at minimal additional cost.
    Adapting the agent to new tasks often only requires high-level adjustments rather than extensive code rewrites.
    This reshapes the economics of every aforementioned challenge:
    expert effort is spent once at the agent layer rather than repeatedly for each case; a single validated agent can extend scarce expertise wherever it is needed; system-specific solutions can be generated by the agent rather than handcrafted; and multiple operating agents meet volumes and timelines that overwhelm human teams.

    \item \textbf{Multi-disciplinary knowledge:} AI agents can integrate knowledge from multiple domains simultaneously, spanning areas such as browser internals, operating systems, compilers, and cryptographic protocols, all of which may be relevant to a single security problem. In contrast, human experts typically specialize deeply in only a subset of these areas, making comprehensive coverage dependent on assembling multiple specialists. By concentrating otherwise fragmented expertise in one place, agents enable cross-domaine reasoning, offering solutions that are not confined to a single discipline, and make it easier to tackle complex problems that require combining knowledge rarely found in any one individual.
\end{enumerate}

\section{CASE STUDIES}

We present \num{16}~case studies across different domains that have faced the challenges outlined above and demonstrate how they can take advantage of \agenticAI for scalable adoption.
Specifically, we look at:

\begin{itemize}
    \item[$\triangleright$] Systems Security
    \item[$\triangleright$] Software Security
    \item[$\triangleright$] Cyber Threat Intelligence (CTI)
    \item[$\triangleright$] Attack Detection
\end{itemize}

\noindent\Cref{tab:case_studies} summarizes these mechanisms and their limitations, which we elaborate on in subsequent sections.\vspace*{-3mm}

\section{SYSTEMS SECURITY}
\subsection{Optimizing Instrumentation for Security} 
\label{sec:instrumentation}
Many security techniques work by adding code to a target system so that relevant behavior can be observed, controlled, or constrained. Canonical examples include partitioning secrets across trusted execution environments\arxivonly{~\cite{10.1145/3590140.3629116}}, adding harnesses and argument generators for fuzzing~\cite{clements2026rtfuzzer}, placing sanitizer and audit hooks in production code\arxivonly{~\cite{203706}}, and inserting declassification points for information-flow control\arxivonly{~\cite{sabelfeld2003language}}. Optimizing instrumentation runs into all five limitations at once. Effective instrumentation demands substantial effort, since the work is performed by hand on large codebases, as long-running projects such as Firefox sandboxing illustrate\arxivonly{~\cite{203706}}. The relevant expertise is scarce, since correct placement requires deep familiarity with both the target program and the analysis it feeds.  Each codebase has its own structure and invariants, so a strategy that works for one target rarely transfers to another. Some level of automation has been achieved more recently but not to a full extent~\cite{9996963}\arxivonly{\cite{203706}}. The underlying problem is complex, requiring joint reasoning over program semantics, the analysis policy, and the cost of each added probe. And in online settings, instrumentation must be produced quickly enough to keep up with the systems it observes. AI agents have the potential to ingest source code, reason over the analysis goal, and propose where and how to instrument, 
turning instrumentation from a multi-month manual effort into a step that scales with the codebases it~protects.

\subsection{Moving Target Defense \& SW Diversification}
\label{sec:moving-target}

Varying the attack surface of a system so that an exploit prepared against one configuration does not work against another~\cite{jajodia2011moving} is a long-standing approach to increase the cost of attacks.
Software diversification is the leading example: rather than shipping a single binary, the defender produces many variants that differ in memory layout, instruction selection, or system-call interface, so that an exploit tuned for one variant fails on the rest.
Network-layer techniques such as address/port shuffling and platform-rotation schemes follow the same principle.
Four factors prevent these techniques from being widely deployed:
First, the changes must occur faster than an attacker can adapt, often on the order of seconds, which is below what human-driven reconfiguration can achieve.
Second, the work requires expertise in compilers, runtime systems, or network architecture that is concentrated in small communities.
Third, every system has its own performance budget, dependencies, and operational constraints, so a diversification strategy designed for one application or network rarely transfers to another.
Fourth, designing an effective moving target defense is inherently challenging, demanding a careful balance between security benefits, performance overhead, verifiability, and compatibility.
Determining what to vary, how frequently to adapt, and at what cost has yet to admit a tractable algorithmic formulation.
AI agents that can analyze a target system, generate diverse variants, and schedule reconfiguration under operational constraints can make moving target defense practical for systems that rely on static configurations and reactive~patching.

\subsection{Formalization of Security Standards}
\label{sec:formalization-of-sec-policies}
The vast majority of standards and security policies are defined primarily in natural language. From public standards through to internal organizational policies, these documents form the basis of modern computing. Over the past two decades we have witnessed significant advantages in formal verification techniques, enabling verification of systems and protocols (e.g., seL4~\cite{klein2009sel4}, TLS~1.3~\cite{cremers2017comprehensive}). However, several factors have limited widespread adoption of formal analysis techniques. First, the work demands substantial human effort to translate from natural language into the precise machine-checkable language used by these tools. Second, this translation requires familiarity with \emph{both} the source domain (e.g., legal, organizational, technical) and the formal semantics of the target language---a combination that could be exceedingly rare. Third, every translation is task-specific: each pairing of source description and target language brings its own ambiguities and idioms, so a translation pipeline built for one pairing rarely transfers to another. AI agents that can read informal requirements, reason over the target policy language, and produce a candidate translation for review could enable this translation to be carried out at scale to unlock the potential of formal verification.

\subsection{Supply Chain Analysis}
\label{sec:supply-chain-analysis}
Complex software projects typically depend on a diverse ecosystem of components, introducing potential attack vectors through dependencies, build infrastructure, or the human elements supporting the supply chain~\cite{williams2025research}.
To monitor dependencies and facilitate identification of imported vulnerabilities, software bill of materials (SBOMs) are usually compared to vulnerability lists. 
However, currently generated SBOMs still face automation challenges: they may be inaccurate or unsound, containing incorrect dependencies or versions~\cite{williams2025research}. 
AI agents, with their ability to ingest and interpret vast quantities of information, are well suited to this challenge. For example, agents can analyze entire repositories to identify dependencies and their versions, and cross-reference these against existing vulnerability repositories. Furthermore, AI agents may offer broader solutions for strengthening the software supply chain. They may defend against project takeover by systematically analyzing publicly available information about new contributors and continuously examining their commits for anomalous or suspicious behavior. 
In doing so, they reduce the reliance on implicit trust and alleviates the unsustainable cognitive and operational burdens historically placed on open-source maintainers.
AI agents could also identify typo-squatting attacks by flagging suspicious package names, enabling timely intervention and removal. Finally, they could prevent dependency confusion by enforcing and monitoring policies that govern how dependencies are resolved and sourced.

\subsection{Self-Healing Systems}
\label{sec:self-healing}
A long-standing vision in systems design is that of systems that detect compromise, diagnose what went wrong, and repair themselves without human intervention~\cite{kephart2003vision}. Today, most deployed systems apply this idea only in a narrow, reliability-oriented sense: restarting failed processes or shifting load to recover from downtime. Recovering from a security incident is a harder problem, because the system must not only survive the attack but also keep the affected process in a correct and consistent state. A repair that lets a program continue past an exploited vulnerability can be worse than a crash if it leaves the program in an unsafe or inconsistent state, for instance by allowing an attacker past an authorization check\arxivonly{~\cite{locasto2007stem}}. Three factors make security self-healing difficult. First, the response must be applied inline, e.g., while the attack is in progress, before the exploit completes its effect. Second, every program has its own logic, integrity constraints, and set of functions that are safe or unsafe to repair, so a recovery strategy that works for one application does not transfer to another. Third, choosing a correct repair is a hard reasoning problem in itself, since it requires understanding what the program was trying to do, which state the attack has corrupted, and which of the available fixes restores a consistent state rather than merely suppressing the symptom.  AI agents can act fast enough to intervene while the attack is still unfolding and reason about the specific program and the ongoing attack to construct a repair tailored to that situation rather than applying a fixed recovery rule. 

\subsection{Customization of System Security}
\label{sec:system-security}
Systems-level security mechanisms such as SELinux, AppArmor, seccomp, and capability-based sandboxing are designed to support fine-grained, per-system tailoring of access control and isolation policies\arxivonly{~\citep{selinux-jaeger,loscocco2001integrating}}. In practice, the policies shipped with most distributions are either too permissive or too complex to adapt, and operators often disable or bypass these policies rather than tuning them to a specific application or threat model.
Two factors make customization difficult:
First, customization is task and domain specific by construction: each deployment has its own application stack, operational context, and risk posture, so a configuration written for one system rarely transfers to another.
Second, the problem itself is complex, requiring joint reasoning over program semantics, system-call behavior, and policy-language details that classical tools handle only in isolation.
AI agents that can observe application behavior, reason over the target policy language, and produce per-system configurations on demand would address both at once, turning customization from a costly expert exercise into a routine deployment step.
AI agents can help with tasks, such as customization of security mechanisms.

\section{SOFTWARE SECURITY}
\subsection{Vulnerability Discovery}
\label{sec:vulnerability-discovery}

Software vulnerabilities are the facilitators of far-reaching security incidents, either as a direct gateway into a computer system or as part of malware.
Although advanced program analysis and AI have accelerated vulnerability discovery in recent years, human expertise has remained the deciding factor for finding flaws in complex software systems.
These days, successful attacks are often a team effort of highly skilled individuals\arxivonly{~\cite{pwn2own}} specialized in subdomains that eventually merge into attack chains of ever-increasing complexity.
Recent developments, such as Claude Mythos\arxivonly{~\cite{claudemythos}}, demonstrate that \agenticAI has the potential to bring another leap in vulnerability discovery.
These systems (a)~cover a wide range of specialized expertise in specific domains (e.g., kernel, browser, smart contracts, cryptographic protocol, web) and attack/analysis techniques;
(b)~can draw far-reaching connections in both depth \emph{and} breadth across codebases, subdomains, and techniques;
(c)~scale well across software projects even if they operate on rare architectures or assume different trust boundaries.
However, \agenticAI has not only great potential for mere discovery, but also for related mechanisms, such as verifying and assessing the severity of vulnerabilities, prioritizing repair, and patch management.

\subsection{Exploit Generation}
\label{sec:exploit-generation}
Finding out whether a software defect is exploitable is inherently difficult.
In contrast to pattern-based vulnerability discovery that locates candidates of exploitable flaws, fuzz testing (fuzzing) already produces inputs that crash a program at runtime and thus reach the software defect via user-controlled input.
However, a significant gap remains between discovering a vulnerability or crash and producing a working exploit~\cite{avgerinos2011aeg}.
In particular, scaling automatic exploit generation to real-world settings requires reasoning about the targeted software as part of a larger system, interacting with a larger environment.
Each of these abstraction levels imposes individual constraints on the exploit, including those via deployed defensive mechanisms.
This complexity makes exploit development notoriously difficult, labor-intensive and, for a long time, a manual effort.
\AgenticAI can bring highly-specialized domain knowledge, reason about far-reaching dependencies among components, and scale nearly without bound.
Moreover, the asymmetry between exploit generation and verification is advantageous: while constructing an exploit is inherently difficult, validating whether a candidate exploit succeeds is typically inexpensive and unambiguous. This asymmetry enables agents to iterate against an oracle, a property that is not always available in other security tasks.

\subsection{Translation to Memory-Safe Languages}
\label{sec:translation-of-legacy-code}
A large fraction of security-critical infrastructure is written in C and C++, and memory-safety violations in this code continue to account for a substantial share of disclosed vulnerabilities. This has motivated broad calls to migrate to memory-safe languages such as Rust~\cite{emre2021translating}, including dedicated efforts such as DARPA's Translating All C to Rust (TRACTOR) program, which explicitly targets automated, idiomatic C-to-Rust translation. Three factors keep such migration out of reach at realistic scale. First, manual rewriting demands substantial human effort: codebases of interest run to hundreds of thousands or millions of lines, and the work cannot be parallelized without expert review of each piece. Second, producing safe and idiomatic translations requires expertise in both the source and target languages, a combination concentrated in a small community. Third, each codebase carries its own idioms, build systems, and invariants, so a translation strategy that works on one project rarely transfers to another, and existing automatic translators fall back on \texttt{unsafe} Rust constructs that preserve the very safety issues the migration was meant to eliminate. AI agents that can read C source, reason about aliasing and ownership, and produce idiomatic Rust under expert supervision could make memory-safety migration tractable for the codebases that need it most.

\subsection{Domain-Specific Code Optimization} 
\label{sec:domain-optimization}
General-purpose security mechanisms are designed to cover a wide range of use cases, but in any specific deployment, only a small subset of their functionality is actually used.
Studies of widely deployed systems show this margin to be substantial:
across more than two thousand programs on a typical Linux desktop, only about five percent of the standard C library is used on average, the rest contributing nothing to the application, but expanding its attack surface \cite{quach2018debloating}.
Examples of optimization for a specific deployment include software and protocol debloating, stripping unused TLS cipher suites or extensions, and reducing kernel or library functionality to what a given application requires.
However, two factors keep such optimization out of reach in most settings. First, the work requires expertise that is concentrated in a small community, since deciding which features can be safely removed depends on detailed knowledge of both the codebase and the threat landscape it serves. Second, every deployment uses a different subset of functionality and faces a different threat profile, so a debloating result for one application or environment does not transfer to another.
AI agents that can analyze a codebase, infer which functionality each deployment actually uses, and produce a stripped-down build could make per-deployment optimization a standard step in deployment rather than a specialist undertaking this challenging~task.

\subsection{Code IP Protection}
\label{sec:code-ip}
Software encodes knowledge for solving a particular problem and thus represents a vendor's intellectual property (IP). 
Consequently, there is a long-standing history of obfuscation/protection schemes for software.
To a large extent, these techniques originate from the virus scene and aim to prevent third parties from analyzing or identifying the presence of a malicious payload.
As early as the 1990s, polymorphic obfuscation came to light~\cite{vonadvanced}, improving over simple encoding and encryption schemes by mutating the decoder/decrypter for every instance.
Instead of selecting from a predefined set of options, the true goal is metamorphism: transforming code while preserving semantics.
Emulator-based obfuscation~\cite{5207639} that uses randomized instruction-to-bytecode mapping is as close to metamorphism as it has got so far.
Truly morphing program code, however, is inherently difficult as it requires extensive reasoning across the codebase and program semantics.
Agentic AI ``transpiling'' code from one (logic) representation to another, and thus code metamorphism, seems within reach both conceptually and in terms of scalability.
Similarly, an AI agent can reason about program semantics, statistical or structural properties, and the threat model simultaneously, enabling more comprehensive protection schemes beyond individual (parts of) programs.

\section{CYBER THREAT INTELLIGENCE}
\subsection{Attribution and Threat Intelligence}
\label{sec:attribution}

When a security incident occurs, defenders need to understand not just what happened but who is behind it, what their broader campaign looks like, and what they are likely to do next. Attribution and threat intelligence aims to answer these questions by combining technical evidence (malware similarity, infrastructure overlap, code reuse) with non-technical sources such as open-source intelligence, leaked documents, language and cultural artifacts, and geopolitical context. However, this is widely regarded as a challenging task for various reasons\arxivonly{~\cite{rid2015attributing}}. First, producing a single well-supported attribution or intelligence report takes weeks of focused analyst work, since evidence must be gathered from many sources and carefully cross-checked. Second, the expertise required is rare and unusual, since the same analyst often needs technical depth in malware and network forensics together with knowledge of relevant languages, regions, and adversary practices. Third, every campaign is its own case: each adversary has its own infrastructure patterns, tooling habits, and operational preferences, and methods that worked on one campaign rarely transfer cleanly to the next. Fourth, attribution itself is a hard reasoning problem, since defenders must weigh evidence under active deception, where adversaries plant false flags and reuse other groups' tools to mislead investigators. AI agents that can collect and correlate evidence across technical and open-source data, propose candidate attributions with their supporting reasoning, and flag where evidence is weak or contradictory could allow attribution and threat intelligence to be produced for many more incidents at a fraction of the cost.

\subsection{Provenance Analysis for Forensics}
\label{sec:provenance-analysis-forensics}

When a security incident occurs, investigators must reconstruct the sequence of events: identifying the initial process, determining which files were accessed, tracing any data exfiltration, and assessing the overall scope of the compromise. Provenance analysis supports this task by recording causal relationships between system entities—such as processes, files, sockets, and users—in a graph structure. This graph can be traversed backward from a detection point to uncover the root cause, or forward to identify downstream effects~\arxivonly{\cite{king2003backtracking}}~\cite{DBLP:journals/tifs/IrshadCGYLPJKXZ21}.
This capability was a central objective of the DARPA Transparent Computing (TC) program\arxivonly{~\cite{darpa_tc}}, which aimed to deliver fine-grained visibility into system interactions across all layers of software while maintaining low performance overhead. The TC program also sought to support both real-time detection and post hoc forensic analysis of advanced persistent threats (APTs). Although substantial progress has been made through DARPA TC and subsequent work~\arxivonly{\cite{10.1145/3105761,10.1145/3564625.3567997,8619416,9095996}}, challenges remain in scaling these approaches, generalizing them across environments, and enabling widespread adoption.
AI agents can help address these gaps by automating key aspects of provenance analysis: interpreting complex provenance graphs, identifying likely attack paths, and explaining the supporting evidence. This added automation has the potential to make forensic provenance analysis a routine part of incident response, rather than a specialized process reserved for high-priority cases.
Moreover, provenance analysis extends beyond individual systems to encompass entire supply chains. In manufacturing, for example, provenance includes information about a part’s components, origin, design, manufacturer, associated software, production data, and quality control records encompassed in cyber-physical passports~\cite{cpp}. In such settings, AI agents can further enhance automation by enabling coordinated forensic analysis across multiple nodes in the supply chain.

\section{ATTACK DETECTION}
\subsection{Insider Threat Detection}
\label{sec:insider-threat-detection}

Some of the most damaging security incidents come from people who already have legitimate access to a system: malicious insiders who misuse their privileges to steal data or sabotage systems, and external attackers who compromise legitimate accounts. Detecting either case is challenging precisely because the activity intentionally looks like normal behavior. A long line of work on insider threat detection has approached this through behavioral analysis, building a model of how each user typically operates and flags deviations\arxivonly{~\cite{salem2008survey}}. However, four factors make this approach challenging to operationalize at scale. First, building and maintaining accurate behavioral models requires substantial manual effort, since each user's normal behavior shifts over time and must be tuned by analysts who understand the organization. Second, when a real insider attack is in progress, the response must be fast: every additional minute the attacker has access translates into more data exfiltrated or more systems compromised; human review of every flagged event is simply too slow. Third, every organization has its own roles, workflows, and acceptable behaviors, so a detection model trained on one company's users rarely transfers directly to another. Fourth, separating malicious activity from legitimate but unusual behavior requires reasoning over the user's role, the current task, the data being touched, and the wider organizational context, which is why both rule-based systems and learned models continue to produce far more false positives than analysts can review. AI agents that can read user activity, build per-user and per-role models of normal behavior, and explain why a given event is suspicious could allow insider threat detection to operate routinely at the scale of entire organizations.

\subsection{Provenance Analysis \& RT Monitoring}
\label{sec:provenance-analysis-monitoring}
Beyond after-the-fact forensics, provenance data can be leveraged for continuous attack detection by monitoring how processes, files, and systems interact in real time. This perspective naturally extends beyond individual systems to encompass entire supply chains, where events occurring across multiple organizations, suppliers, and production stages form a distributed provenance graph. However, deploying such monitoring at enterprise or supply chain scale remains a significant challenge. Even organizations of modest size generate tens of thousands of alerts each week, far more than analysts can realistically investigate, while the underlying event stream can reach hundreds of millions of records per day~\cite{hassan2019nodoze}. When expanded across interconnected supply chain nodes, this volume and complexity grow even further.

Three factors keep real-time provenance monitoring difficult to operate at this scale. First, finding the relevant subgraph in a continuous high-volume stream takes substantial manual effort, and analysts who can do it well are rare, so most alerts go uninvestigated. Second, the value of an alert drops quickly: a real attack that is flagged hours after it begins is much harder to contain than one caught while it is still in its early stages, and human review of a queue with thousands of alerts cannot keep up. Third, deciding whether a sequence of events is benign or malicious requires reasoning at once over normal system behavior, the structure of the graph, and the specific patterns that distinguish real attacks from background noise. AI agents that can read live provenance streams, identify the small fraction of subgraphs that need attention, and explain their reasoning could allow real-time provenance monitoring to keep pace with the data that defenders already collect but cannot yet act on.

\subsection{Adaptive Honeypots}
\label{sec:dynamic-honeypots}
Honeypots are electronic bait that lure attackers into interacting with it, allowing to assess and analyze their tactics, techniques, and procedures~(TTPs)~\cite{4627312}.
There are two prevalent strategies: (a)~passive honeypots and (b)~active honeypots.
The former is deployed (e.g.,~in a network's unmapped IP~range) with the sole purpose of passively monitoring any incoming interaction and increasing/maintaining engagement.
The latter actively set out interact with a system (e.g.,~a potentially harmful web page) imitating a ``victim.''
For both, \agenticAI can be beneficial, allowing for the creation of significantly more complex bait.
In the passive setting, AI agents can act as a collective of honeypots imitating a complex network structure, where interaction can happen in accordance to an adversary's expectation.
Additionally, \agenticAI can realistically populate observable communication (e.g., network traffic).
In contrast to ``traditional'' setups, honeypots based on AI agents can adapt to adversaries on an individual basis, giving rise to effective analysis of even lateral movement in sophisticated and highly specific attacks.
For active honeypots, in turn, the ability to adapt to the potential attacker is equally essential to extract as much information as possible.
For instance, multiple agents can effectively probe and bypass environment fingerprinting \mbox{(as used for web-based exploit kits)---even} for environments and attack prerequisites not known upfront.

\section{LIMITATIONS \& OPEN QUESTIONS}
As discussed in the previous section, AI agents have the potential to revive previously impractical or abandoned security strategies, enabling solutions to problems that were once considered intractable. However, agents are not a panacea and introduce their own set of challenges. For example, their inherently stochastic nature necessitates verification\arxivonly{~\cite{meng2022adversarial}} and inspection\arxivonly{~\cite{hussain2025transparency}} to ensure reliability and correctness of outputs.
An additional challenge lies in evaluation. Benchmarking agent performance is difficult: once a metric becomes a target for optimization, the benchmark may defeat its purpose\arxivonly{~\cite{goodhart1984problems}}. Recent work has therefore proposed, for example, a comparative approach between models across benchmarks\arxivonly{~\cite{zhangbenchmark}}. However, a fundamental question remains: how to evaluate agents' performance when they surpass human capabilities?
These difficulties are compounded when considering the security of the AI agents themselves\arxivonly{~\cite{tramer2020adaptive}}. More precisely, agents inherit the vulnerabilities of the underlying AI systems\arxivonly{~\cite{biggio2018wild}}, while simultaneously amplifying risk through autonomy and privileged access, if misconfigured\arxivonly{~\cite{christodorescu2026agents}}. Securing such AI systems is inherently difficult and these vulnerabilities affect AI agents' transparency\arxivonly{~\cite{vadillo2025adversarial}}.
Given the interplay between AI-specific vulnerabilities\arxivonly{~\cite{tramer2020adaptive,biggio2018wild}} and system-level security concerns\arxivonly{~\cite{christodorescu2026agents}}, adopting \agenticAI approaches requires caution---especially for security-critical settings.
Agents risk amplifying existing weaknesses or obscuring system behavior. A principled approach that explicitly accounts for these  challenges is therefore~essential.

\section{CONCLUSION}
\AgenticAI will lead to significant changes in all areas of technology.
In cybersecurity specifically, an exciting prospect is how \agenticAI can be used to overcome challenges that have previously prevented various security techniques from reaching their full potential if open questions and limitations are overcome.
The case studies discussed above have been chosen to illustrate this potential, but this list is by no means exhaustive.
We therefore encourage the community to reexamine past assumptions and revisit past security techniques with the assistance of \agenticAI.

\section{ACKNOWLEDGMENTS}
We thank {Christian Schroeder de Witt} for contributing to the initial discussions during Dagstuhl Perspectives Workshop 26162: \emph{``Autonomous AI Agents in Computer Security,''} that lead to this work.

This material is supported in part by the U.S. Department of Energy’s Office of Energy Efficiency and Renewable Energy (EERE) under the Advanced Materials and Manufacturing Technologies Office, Award Number DE-EE0009046, the DARPA under agreement number 885000, the NSF CCF-FMiTF-1836978, and ONR N00014-21-1-2492.
The views expressed herein do not necessarily represent the views of the U.S. Department of Energy or the United States Government.

Moreover, we gratefully acknowledge support by the Austrian Federal Ministry of Economy, Energy and Tourism, the National Foundation for Research, Technology and Development and the Christian Doppler Research Association, the SBA Research (SBA-K1 NGC), a COMET Center funded by BMIMI, BMWET, the federal state of Vienna, managed by FFG, and the Helmholtz Association (HGF) within topic ''46.23 Engineering Secure Systems.''

\def\refname{REFERENCES}

\bibliographystyle{plain}
\bibliography{lit}

@misc{claudemythos,
author={Nicholas Carlini and Newton Cheng and Keane Lucas and Michael Moore and Milad Nasr and Vinay Prabhushankar and  Winnie Xiao
 and Hakeem Angulu and Evyatar Ben Asher, Jackie Bow and Keir Bradwell  and Ben Buchanan  and David Forsythe  and Daniel Freeman and Alex Gaynor and Xinyang Ge  and Logan Graham and Kyla Guru  and Hasnain Lakhani  and Matt McNiece  and Mojtaba Mehrara  and Renee Nichol  and Adnan Pirzada  and Sophia Porter and Andreas Terzis  and Kevin Troy},
title={Assessing Claude Mythos Preview’s cybersecurity capabilities},
howpublished={\url{https://red.anthropic.com/2026/mythos-preview/}}
}

@INPROCEEDINGS{4627312,
  author={Watson, David and Riden, Jamie},
  booktitle={2008 WOMBAT Workshop on Information Security Threats Data Collection and Sharing}, 
  title={The Honeynet Project: Data Collection Tools, Infrastructure, Archives and Analysis}, 
  year={2008},
  volume={},
  number={},
  pages={24-30},
  doi={10.1109/WISTDCS.2008.11}}

@misc{pwn2own,
author={Baran, Guru},
title={{Hackers Exploited 73 0-Day Vulnerabilities and Earned \$1,024,750}},
howpublished={\url{https://cybersecuritynews.com/73-unique-0-day-vulnerabilities-pwn2own/}}
}

@article{vonadvanced,
  title={Advanced Code Evolution Techniques and Computer Virus Generator Kits},
  author={von Neumann, John}
}

@INPROCEEDINGS{5207639,
  author={Sharif, Monirul and Lanzi, Andrea and Giffin, Jonathon and Lee, Wenke},
  booktitle={2009 30th IEEE Symposium on Security and Privacy}, 
  title={Automatic Reverse Engineering of Malware Emulators}, 
  year={2009},
  volume={},
  number={},
  pages={94-109},
  doi={10.1109/SP.2009.27}}

@inproceedings{selinux-jaeger,
title="Analyzing Integrity Protection in the SELinux Example Policy",
author="Trent Jaeger and	Reiner Sailer and 	Xiaolan Zhang",
booktitle="Usenix Security",
year="2003"
}

@misc{cpp,
author={Howard D. Grimes and Gabriela F. Ciocarlie and Robert J. Butler and Wayne E. Austad},
title={{Microelectronics Offer Case Study for Securing Defense-Critical Supply Chains}},
howpublished={\url{https://www.afcea.org/signal-media/cyber-edge/microelectronics-offer-case-study-securing-defense-critical-supply-chains}}
}

@INPROCEEDINGS{9095996,
  author={Sahabandu, Dinuka and Allen, Joey and Moothedath, Shana and Bushnell, Linda and Lee, Wenke and Poovendran, Radha},
  booktitle={2020 ACM/IEEE 11th International Conference on Cyber-Physical Systems (ICCPS)}, 
  title={Quickest Detection of Advanced Persistent Threats: A Semi-Markov Game Approach}, 
  year={2020},
  volume={},
  number={},
  pages={9-19},
  doi={10.1109/ICCPS48487.2020.00009}}

@INPROCEEDINGS{8619416,
  author={Sahabandu, Dinuka and Xiao, Baicen and Clark, Andrew and Lee, Sangho and Lee, Wenke and Poovendran, Radha},
  booktitle={2018 IEEE Conference on Decision and Control (CDC)}, 
  title={DIFT Games: Dynamic Information Flow Tracking Games for Advanced Persistent Threats}, 
  year={2018},
  volume={},
  number={},
  pages={1136-1143},
  doi={10.1109/CDC.2018.8619416}}

@inproceedings{10.1145/3564625.3567997,
author = {Liu, Yushan and Shu, Xiaokui and Sun, Yixin and Jang, Jiyong and Mittal, Prateek},
title = {RAPID: Real-Time Alert Investigation with Context-aware Prioritization for Efficient Threat Discovery},
year = {2022},
isbn = {9781450397599},
publisher = {Association for Computing Machinery},
address = {New York, NY, USA},
url = {https://doi.org/10.1145/3564625.3567997},
doi = {10.1145/3564625.3567997},
booktitle = {Proceedings of the 38th Annual Computer Security Applications Conference},
pages = {827–840},
numpages = {14},
location = {Austin, TX, USA},
series = {ACSAC '22}
}

@article{10.1145/3105761,
author = {Shu, Xiaokui and Yao, Danfeng (Daphne) and Ramakrishnan, Naren and Jaeger, Trent},
title = {Long-Span Program Behavior Modeling and Attack Detection},
year = {2017},
issue_date = {November 2017},
publisher = {Association for Computing Machinery},
address = {New York, NY, USA},
volume = {20},
number = {4},
issn = {2471-2566},
url = {https://doi.org/10.1145/3105761},
doi = {10.1145/3105761},
journal = {ACM Trans. Priv. Secur.},
month = sep,
articleno = {12},
numpages = {28}
}

@misc{darpa_tc,
author={DARPA},
title={{Transparent Computing}},
howpublished={\url{https://www.darpa.mil/research/programs/transparent-computing}}
}

@article{DBLP:journals/tifs/IrshadCGYLPJKXZ21,
  author       = {Hassaan Irshad and
                  Gabriela F. Ciocarlie and
                  Ashish Gehani and
                  Vinod Yegneswaran and
                  Kyu Hyung Lee and
                  Jignesh M. Patel and
                  Somesh Jha and
                  Yonghwi Kwon and
                  Dongyan Xu and
                  Xiangyu Zhang},
  title        = {{TRACE:} Enterprise-Wide Provenance Tracking for Real-Time {APT} Detection},
  journal      = {{IEEE} Trans. Inf. Forensics Secur.},
  volume       = {16},
  pages        = {4363--4376},
  year         = {2021},
  url          = {https://doi.org/10.1109/TIFS.2021.3098977},
  doi          = {10.1109/TIFS.2021.3098977},
  bibsource    = {dblp computer science bibliography, https://dblp.org}
}

@inproceedings {203706,
author = {Shiqing Ma and Juan Zhai and Fei Wang and Kyu Hyung Lee and Xiangyu Zhang and Dongyan Xu},
title = {{MPI}: Multiple Perspective Attack Investigation with Semantic Aware Execution Partitioning},
booktitle = {26th USENIX Security Symposium (USENIX Security 17)},
year = {2017},
isbn = {978-1-931971-40-9},
address = {Vancouver, BC},
pages = {1111--1128},
url = {https://www.usenix.org/conference/usenixsecurity17/technical-sessions/presentation/ma},
publisher = {USENIX Association},
month = aug
}

@inproceedings{clements2026rtfuzzer,
  title     = {RT-Fuzzer: Task Driven Fuzzing of Real Time Operating System Firmware},
  author    = {Clements, Abraham and Gomez Rivera, Abel and Liu, Richard Jiayang and Levchenko, Kirill and Kennell, Rick and Ciocarlie, Gabriela},
  booktitle = {NDSS Workshop on Binary Analysis Research (BAR)},
  year      = {2026},
  url       = {https://www.ndss-symposium.org/wp-content/uploads/bar2026-50.pdf}
}

@inproceedings{10.1145/3590140.3629116,
author = {Yuhala, Peterson and Felber, Pascal and Guiroux, Hugo and Lozi, Jean-Pierre and Tchana, Alain and Schiavoni, Valerio and Thomas, Ga\"{e}l},
title = {SecV: Secure Code Partitioning via Multi-Language Secure Values},
year = {2023},
isbn = {9798400701771},
publisher = {Association for Computing Machinery},
address = {New York, NY, USA},
url = {https://doi.org/10.1145/3590140.3629116},
doi = {10.1145/3590140.3629116},
booktitle = {Proceedings of the 24th International Middleware Conference},
pages = {207–219},
numpages = {13},
location = {Bologna, Italy},
series = {Middleware '23}
}

@ARTICLE{9996963,
  author={Alhanahnah, Mohannad and Ma, Shiqing and Gehani, Ashish and Ciocarlie, Gabriela F. and Yegneswaran, Vinod and Jha, Somesh and Zhang, Xiangyu},
  journal={IEEE Transactions on Software Engineering}, 
  title={autoMPI: Automated Multiple Perspective Attack Investigation With Semantics Aware Execution Partitioning}, 
  year={2023},
  volume={49},
  number={4},
  pages={2761-2775},
  doi={10.1109/TSE.2022.3231242}}

@article{vadillo2025adversarial,
  title={Adversarial attacks in explainable machine learning: A survey of threats against models and humans},
  author={Vadillo, Jon and Santana, Roberto and Lozano, Jose A},
  journal={Wiley Interdisciplinary Reviews: Data Mining and Knowledge Discovery},
  volume={15},
  number={1},
  pages={e1567},
  year={2025},
  publisher={Wiley Online Library}
}

@article{hussain2025transparency,
  title={Transparency and accountability: unpacking the real problems of explainable AI},
  author={Hussain, Afzal and Hussain, Ashfaq},
  journal={AI \& SOCIETY},
  volume={40},
  number={7},
  pages={5587--5588},
  year={2025},
  publisher={Springer}
}

@article{meng2022adversarial,
  title={Adversarial robustness of deep neural networks: A survey from a formal verification perspective},
  author={Meng, Mark Huasong and Bai, Guangdong and Teo, Sin Gee and Hou, Zhe and Xiao, Yan and Lin, Yun and Dong, Jin Song},
  journal={IEEE Transactions on Dependable and Secure Computing},
  year={2022},
  publisher={IEEE}
}

@article{williams2025research,
  title={Research directions in software supply chain security},
  author={Williams, Laurie and Benedetti, Giacomo and Hamer, Sivana and Paramitha, Ranindya and Rahman, Imranur and Tamanna, Mahzabin and Tystahl, Greg and Zahan, Nusrat and Morrison, Patrick and Acar, Yasemin and others},
  journal={ACM Transactions on Software Engineering and Methodology},
  volume={34},
  number={5},
  pages={1--38},
  year={2025},
  publisher={ACM New York, NY}
}

@article{tramer2020adaptive,
  title={On adaptive attacks to adversarial example defenses},
  author={Tramer, Florian and Carlini, Nicholas and Brendel, Wieland and Madry, Aleksander},
  journal={Advances in neural information processing systems},
  volume={33},
  pages={1633--1645},
  year={2020}
}

@techreport{cobos2024review,
  author      = {Cobos, E. and Cakir, Selcen},
  title       = {A Review of the Economic Costs of Cyber Incidents},
  institution = {World Bank Group},
  year        = {2024},
  type        = {Report},
}

@incollection{goodhart1984problems,
  title={Problems of monetary management: the UK experience},
  author={Goodhart, Charles AE},
  booktitle={Monetary theory and practice: The UK experience},
  pages={91--121},
  year={1984},
  publisher={Springer}
}

@article{acharya2025agentic,
  title={Agentic AI: Autonomous intelligence for complex goals—A comprehensive survey},
  author={Acharya, Deepak Bhaskar and Kuppan, Karthigeyan and Divya, B},
  journal={IEEe Access},
  volume={13},
  pages={18912--18936},
  year={2025},
  publisher={IEEE}
}

@inproceedings{zhou2024webarena,
  title={Webarena: A realistic web environment for building autonomous agents},
  author={Zhou, Shuyan and Xu, Frank F and Zhu, Hao and Zhou, Xuhui and Lo, Robert and Sridhar, Abishek and Cheng, Xianyi and Ou, Tianyue and Bisk, Yonatan and Fried, Daniel and others},
  booktitle={International Conference on Learning Representations},
  volume={2024},
  pages={15585--15606},
  year={2024}
}

@article{jiang2026survey,
  title={A survey on large language models for code generation},
  author={Jiang, Juyong and Wang, Fan and Shen, Jiasi and Kim, Sungju and Kim, Sunghun},
  journal={ACM Transactions on Software Engineering and Methodology},
  volume={35},
  number={2},
  pages={1--72},
  year={2026},
  publisher={ACM New York, NY}
}

@article{radford2019language,
  title={Language models are unsupervised multitask learners},
  author={Radford, Alec and Wu, Jeffrey and Child, Rewon and Luan, David and Amodei, Dario and Sutskever, Ilya and others},
  journal={OpenAI blog},
  year={2019}
}

@inproceedings{zhangbenchmark,
  title={How Benchmark Prediction from Fewer Data Misses the Mark},
  author={Zhang, Guanhua and Dorner, Florian E. and Hardt, Moritz},
  booktitle={The Thirty-ninth Annual Conference on Neural Information Processing Systems},
  year={2025}
}

@article{christodorescu2026agents,
      title={Agent Security is a Systems Problem}, 
      author={Mihai Christodorescu and Earlence Fernandes and Ashish Hooda and Somesh Jha and Johann Rehberger and Kamalika Chaudhuri and Xiaohan Fu and Khawaja Shams and Guy Amir and Jihye Choi and Sarthak Choudhary and Nils Palumbo and Andrey Labunets and Nishit V. Pandya},
      year={2026},
      journal={arXiv preprint arxiv.org:2605.18991}
}

@article{sabelfeld2003language,
  author={Sabelfeld, A. and Myers, A.C.},
  journal={IEEE Journal on Selected Areas in Communications}, 
  title={Language-based information-flow security}, 
  year={2003},
  volume={21},
  number={1},
  pages={5-19},
  doi={10.1109/JSAC.2002.806121}}

@article{emre2021translating,
    author = {Emre, Mehmet and Schroeder, Ryan and Dewey, Kyle and Hardekopf, Ben},
    title = {Translating C to safer Rust},
    year = {2021},
    issue_date = {October 2021},
    publisher = {Association for Computing Machinery},
    address = {New York, NY, USA},
    volume = {5},
    number = {OOPSLA},
    url = {https://doi.org/10.1145/3485498},
    doi = {10.1145/3485498},
    journal = {Proc. ACM Program. Lang.},
    month = oct,
    articleno = {121},
    numpages = {29}
}

@inproceedings {loscocco2001integrating,
    author = {Peter Loscocco and Stephen Smalley},
    title = {Integrating Flexible Support for Security Policies into the Linux Operating System},
    booktitle = {2001 USENIX Annual Technical Conference (USENIX ATC 01)},
    year = {2001},
    address = {Boston, MA},
    url = {https://www.usenix.org/conference/2001-usenix-annual-technical-conference/integrating-flexible-support-security-policies},
    publisher = {USENIX Association},
    month = jun
}

@inproceedings {quach2018debloating,
    author = {Anh Quach and Aravind Prakash and Lok Yan},
    title = {Debloating Software through {Piece-Wise} Compilation and Loading},
    booktitle = {27th USENIX Security Symposium (USENIX Security 18)},
    year = {2018},
    isbn = {978-1-939133-04-5},
    address = {Baltimore, MD},
    pages = {869--886},
    url = {https://www.usenix.org/conference/usenixsecurity18/presentation/quach},
    publisher = {USENIX Association},
    month = aug
}

@inproceedings{biggio2018wild,
  title={Wild patterns: Ten years after the rise of adversarial machine learning},
  author={Biggio, Battista and Roli, Fabio},
  booktitle={Proceedings of the 2018 ACM SIGSAC conference on computer and communications security},
  pages={2154--2156},
  year={2018}
}

@article{kephart2003vision,
  author={Kephart, J.O. and Chess, D.M.},
  journal={Computer}, 
  title={The vision of autonomic computing}, 
  year={2003},
  volume={36},
  number={1},
  pages={41-50},
  doi={10.1109/MC.2003.1160055}
  }

@book{jajodia2011moving,
  title = {Moving {{Target Defense}}: {{Creating Asymmetric Uncertainty}} for {{Cyber Threats}}},
  shorttitle = {Moving {{Target Defense}}},
  editor = {Jajodia, Sushil and Ghosh, Anup K. and Swarup, Vipin and Wang, Cliff and Wang, X. Sean},
  year = 2011,
  series = {Advances in {{Information Security}}},
  volume = {54},
  publisher = {Springer New York},
  address = {New York, NY},
  doi = {10.1007/978-1-4614-0977-9},
  urldate = {2026-04-28},
  copyright = {https://www.springernature.com/gp/researchers/text-and-data-mining},
  isbn = {978-1-4614-0976-2 978-1-4614-0977-9},
  langid = {english}
}

@inproceedings{avgerinos2011aeg,
  author       = {Thanassis Avgerinos and
                  Sang Kil Cha and
                  Brent Lim Tze Hao and
                  David Brumley},
  title        = {{AEG:} Automatic Exploit Generation},
  booktitle    = {Proceedings of the Network and Distributed System Security Symposium,
                  {NDSS} 2011, San Diego, California, USA, 6th February - 9th February
                  2011},
  publisher    = {The Internet Society},
  year         = {2011},
}

@inproceedings{king2003backtracking,
    author = {King, Samuel T. and Chen, Peter M.},
    title = {Backtracking intrusions},
    year = {2003},
    isbn = {1581137575},
    publisher = {Association for Computing Machinery},
    address = {New York, NY, USA},
    url = {https://doi.org/10.1145/945445.945467},
    doi = {10.1145/945445.945467},
    booktitle = {Proceedings of the Nineteenth ACM Symposium on Operating Systems Principles},
    pages = {223–236},
    numpages = {14},
    location = {Bolton Landing, NY, USA},
    series = {SOSP '03}
}

@inproceedings{hassan2019nodoze,
  title = {{{NoDoze}}: {{Combatting Threat Alert Fatigue}} with {{Automated Provenance Triage}}},
  shorttitle = {{{NoDoze}}},
  booktitle = {Proceedings 2019 {{Network}} and {{Distributed System Security Symposium}}},
  author = {Hassan, Wajih Ul and Guo, Shengjian and Li, Ding and Chen, Zhengzhang and Jee, Kangkook and Li, Zhichun and Bates, Adam},
  year = 2019,
  publisher = {Internet Society},
  address = {San Diego, CA},
  doi = {10.14722/ndss.2019.23349},
  urldate = {2026-04-28},
  isbn = {978-1-891562-55-6},
  langid = {english},
}

@article{rid2015attributing,
    author = {Thomas Rid and Ben Buchanan},
    title = {Attributing Cyber Attacks},
    journal = {Journal of Strategic Studies},
    volume = {38},
    number = {1-2},
    pages = {4--37},
    year = {2015},
    publisher = {Routledge},
    doi = {10.1080/01402390.2014.977382}
}

@inbook{salem2008survey,
    author="Salem, Malek Ben
    and Hershkop, Shlomo
    and Stolfo, Salvatore J.",
    editor="Stolfo, Salvatore J.
    and Bellovin, Steven M.
    and Keromytis, Angelos D.
    and Hershkop, Shlomo
    and Smith, Sean W.
    and Sinclair, Sara",
    title="A Survey of Insider Attack Detection Research",
    bookTitle="Insider Attack and Cyber Security: Beyond the Hacker",
    year="2008",
    publisher="Springer US",
    address="Boston, MA",
    pages="69--90",
    isbn="978-0-387-77322-3",
    doi="10.1007/978-0-387-77322-3_5",
    url="https://doi.org/10.1007/978-0-387-77322-3_5"
}

@inproceedings{klein2009sel4,
  title={seL4: Formal verification of an OS kernel},
  author={Klein, Gerwin and Elphinstone, Kevin and Heiser, Gernot and Andronick, June and Cock, David and Derrin, Philip and Elkaduwe, Dhammika and Engelhardt, Kai and Kolanski, Rafal and Norrish, Michael and others},
  booktitle={Proceedings of the ACM SIGOPS 22nd symposium on Operating systems principles},
  pages={207--220},
  year={2009}
}

@inproceedings{cremers2017comprehensive,
  title={A comprehensive symbolic analysis of TLS 1.3},
  author={Cremers, Cas and Horvat, Marko and Hoyland, Jonathan and Scott, Sam and Van Der Merwe, Thyla},
  booktitle={Proceedings of the 2017 ACM SIGSAC conference on computer and communications security},
  pages={1773--1788},
  year={2017}
}

@inproceedings{locasto2007stem,
author = {Locasto, Michael E. and Stavrou, Angelos and Cretu, Gabriela F. and Keromytis, Angelos D.},
title = {From STEM to SEAD: speculative execution for automated defense},
year = {2007},
isbn = {9998888776},
publisher = {USENIX Association},
address = {USA},
booktitle = {2007 USENIX Annual Technical Conference on Proceedings of the USENIX Annual Technical Conference},
articleno = {17},
numpages = {14},
location = {Santa Clara, CA},
series = {ATC'07}
}

\end{document}